\documentclass[twocolumn,showpacs,showkeys,superscriptaddress]{revtex4}

\usepackage{amsmath}
\usepackage{bbold}
\usepackage{amsfonts}
\usepackage{amssymb}
\usepackage{pbsi}
\usepackage[T1]{fontenc}
\usepackage{xcolor}
\usepackage{graphicx}
\usepackage{subfig}
\usepackage{float}
\usepackage{color}
\usepackage{hyperref}
\hypersetup{colorlinks = true,
	linkcolor = red,
	anchorcolor = blue,
	citecolor = blue,
	filecolor = blue,
	urlcolor = blue}

\begin{document}

\title{Exact solutions to the Telegraph equation in terms of Airy functions}

\author{Felipe A. Asenjo}
\email{felipe.asenjo@uai.cl}
\affiliation{Facultad de Ingenier\'ia y Ciencias,
Universidad Adolfo Ib\'a\~nez, Santiago 7491169, Chile.}
\author{Sergio A. Hojman}
\email{sergio.hojman@uai.cl}
\affiliation{Departamento de Ciencias, Facultad de Artes Liberales,
Universidad Adolfo Ib\'a\~nez, Santiago 7491169, Chile.}
\affiliation{Departamento de F\'{\i}sica, Facultad de Ciencias, Universidad de Chile,
Santiago 7800003, Chile.}
\author{Braulio M. Villegas-Mart\'inez}
\email{bvillegas@inaoep.mx}
\affiliation{Instituto Nacional de Astrof\'isica, \'Optica y Electr\'onica. Calle Luis Enrique Erro No. 1,
Santa Mar\'ia Tonantzintla, Puebla 72840, Mexico.}
\author{H\'ector M. Moya-Cessa}
\email{hmmc@inaoep.mx}
\affiliation{Instituto Nacional de Astrof\'isica, \'Optica y Electr\'onica. Calle Luis Enrique Erro No. 1,
Santa Mar\'ia Tonantzintla, Puebla 72840, Mexico.}
\author{Francisco Soto-Eguibar}
\email{feguibar@inaoep.mx}\affiliation{Instituto Nacional de Astrof\'isica, \'Optica y Electr\'onica. Calle Luis Enrique Erro No. 1,
Santa Mar\'ia Tonantzintla, Puebla 72840, Mexico.}

\date{\today}

\begin{abstract}
Two exact different solutions to the Telegraph equation in three-dimensional space are obtained in terms of Airy functions. As a result, these solutions unveil a distinctive propagation pattern along a coordinate that resembles a speed-cone-like coordinate of the system. This unique characteristic leads to effective Schr\"odinger-like equations, amenable to exact solutions through Airy functions. 
\end{abstract}

\pacs{}

\keywords{}

\maketitle

\section{Introduction}

Berry and Balazs' groundbreaking work \cite{berry},  which demonstrates non-diffracting solutions for quantum mechanical free-particles through Airy functions, has gained substantial attention in recent years from both physics and engineering communities. Notably,  Airy wavepackets have been identified as versatile solutions across a wide spectrum of wave phenomena, ranging from from optics to fluid dynamics, acoustics, thermal conduction (see Refs.~\cite{bookolivier,neomi,Jiang,abdo,bouch,esat,panag,moya,Baumgartl,Nikolaos,chong, Kaminer,water,chencho,zhao,asenjohojmanT}, and references therein), among others. Remarkably, these solutions have been identified in the context of both Schr\"odinger-like equations and hyperbolic space-time equations, thereby highlighting their fundamental significance within a wide range of mathematical and physical frameworks. 

In light of these developments, this work delves into the study of non-diffracting solutions for the well-established Telegraph equation \cite{kraus,hayt,marshall}. Although several well-documented solutions for this equation exist in diverse domains (see for example Refs.~\cite{Dehghan,Gold,ChenChen,BiazarM,kumar,Gombosi,fhuang,Blumana,DasaK,Mostafa,Vineet,Beghin2,Beghin,Turbin,Borodin}), including those in terms of Bessel functions \cite{Borisov},  we show here that Airy functions also emerge as a three-dimensional solution for propagating wavepackets of the general Telegraph equation, written as
\begin{equation}
\frac{\partial^2 u}{\partial t^2}+\alpha  \frac{\partial u}{\partial t}-c^2 \nabla^2 u+ \beta u=0\, ,
\label{telegraphic}
\end{equation}
for some physical dynamical quantity $u$ evolving in time $t$ and three-dimensional space with Cartesian coordinates $x,y,z$, and where ${\nabla}^2$ is the Laplacian. Here, $c$ is a constant that can be identified with the characteristic speed of the system, and $\alpha$ and $\beta$ are two arbitrary constants with units of frequency and square frequency, respectively. 

The Telegraph equation \eqref{telegraphic} plays a major role in several physics fields of study.  It is enough to choose the parameters $\alpha$, $\beta$ and $c$ properly to switch from one physical system to another. Depending on the values of such constants (which could be either real or complex), we can describe several different phenomena. For example, as originally formulated, the  Telegraph equation describes the propagation of an electric signal along a coaxial transmission line (such as RLC circuits), when $\alpha=G/C+R/L$, $\beta=RG/LC$, and $c=1/\sqrt{LC}$, where $R$, $L$, $G$ and $C$ are 
the resistance, the inductance, the conductance and the capacitance, respectively \cite{fabrizio,Puria}. We can also model electromagnetic field propagation in a conducting medium when $c=1/\sqrt{\mu\epsilon}$ represents the speed of light in such medium (with the permittivity $\epsilon$ and permeability $\mu$), 
$\alpha=\sigma/\epsilon$ (where $\sigma$ is the medium conductivity), and $\beta=0$ \cite{fabrizio,Thompson}. Besides, this equation is used to study extended diffusive systems  and random walks process when $\alpha=1/T$ and $\beta=0$, where $T$ is a characteristic time \cite{Litvinenko,Weiss}.
Furthermore, the Telegraph equation describes extended or relativistic quantum mechanical systems when $c$ is the speed of light,
$\alpha=\pm i 2 m c^2/\hbar$ (where $\hbar$ is the reduced Planck constant and $m$ is the mass of the particle) and $\beta=0, \, m^2 c^4/\hbar^2, \, 2 m c^2$ or $V/\hbar^2$, depending on the quantum system to be described (where $V$ is a potential)
\cite{Arbab,Sancho,kostin}. Even more, Eq.~\eqref{telegraphic} has been used in biomedicine for optical imaging,  when $\alpha=3+2\mu_s$ and $\beta=\mu_s(3+\mu_s)$, being $\mu_s$ a normalized inverse scattering length of the medium \cite{Weiss, Arridge}.

All of the above examples show that knowing new solutions to the (old) Telegraph equation \eqref{telegraphic} can bring new insights into several different physical phenomena. That is the purpose of this work.

\section{Airy solutions}

In order to find a solution in terms of Airy functions, let us start  assuming  a solution of the form
\begin{equation}
    u(t,x,y,z)=f(\zeta,y,z)\exp\left(\frac{\gamma}{c}\eta+\frac{\beta+\alpha\gamma}{c^2\lambda} \zeta\right)\, ,
    \label{asn1}
\end{equation}
where $\zeta=x-ct$ and $\eta=x+c t$ are  characteristic speed-cone-like coordinates \cite{imb}, $\gamma$ is an arbitrary constant, and
$\lambda\equiv (4\gamma+\alpha)/c$. This ansatz is invoked to obtain a Schr\"odinger-like equation (Fourier-like heat equation). 
Thus, substituting \eqref{asn1} in the Telegraph equation \eqref{telegraphic}, we find
\begin{equation}
    \lambda \frac{\partial f}{\partial\zeta}=-\left(\frac{\partial^2}{\partial y^2}+\frac{\partial^2}{\partial z^2}\right)f\, .
    \label{schro2d}
\end{equation}
 
From this point on, we can straightforwardly solve the above equation in terms of Airy functions in two different ways. They will be discussed separately.

\begin{widetext}
  
\subsection{First kind of solution}

We can look for a solution in which each spatial dynamic is decoupled from the other one. This is achieved by considering in Eq.~\eqref{schro2d} the ansatz    
\begin{equation}
f(\zeta,y,z)=f_y(\zeta,y)f_z(\zeta,z)\, ,
\end{equation}
such that  we find the simplest solution 
\begin{equation}\label{0050}
     \lambda \frac{\partial f_j}{\partial\zeta}=- \frac{\partial^2 f_j}{\partial j^2}\, ,
\end{equation}
for $j\equiv y,z$, where the separation constants are taken to vanish. These previous equations correspond to one-dimensional spatial free Schr\"odinger-like equations. Therefore, following Berry and Balazs \cite{berry}, we readily find that one possible solution for each of them is in terms of an Airy function, spell it as
\begin{eqnarray}
     f_j(\zeta,j)={\mbox{Ai}}\left(\left(\frac{a_j \lambda^2}{2} \right)^{1/3}\left(j+\frac{a_j}{2}\zeta^2 \right)\right)
     \exp\left(-\frac{a_j\lambda}{2}\zeta \left(j+\frac{a_j}{3}\zeta^2 \right) \right)\, ,  
     \label{airysolution}
\end{eqnarray}
where ${\mbox{Ai}}$ is the Airy function, and $a_j$ is an arbitrary constant that usually is interpreted as the acceleration induced on the maximum intensity lobe of the Airy function in the $j$-direction. 
Thus, the most general solution of Eq.~\eqref{telegraphic} of this kind (in Cartesian coordinates) is
\begin{eqnarray}
    u(t,x,y,z)&=&{\mbox{Ai}}\left(\left(\frac{a_y \lambda^2}{2} \right)^{1/3}\left(y+\frac{a_y}{2}\zeta^2 \right)\right){\mbox{Ai}}\left(\left(\frac{a_z \lambda^2}{2} \right)^{1/3}\left(z+\frac{a_z}{2}\zeta^2 \right)\right)\nonumber\\
&&\times  \exp\left(-\frac{\lambda}{2}\zeta \left(a_y y+a_z z+\frac{a_y^2+a_z^2}{3}\zeta^2 \right) +\frac{\gamma}{c}\eta+\frac{\beta+\alpha\gamma}{c^2\lambda} \zeta\right)\, .
    \label{asnG}
\end{eqnarray}

\subsection{Second kind of solution}

Instead of having independent dynamics on the spatial directions, we can construct a single solution of Eq.~\eqref{schro2d} that present a single accelerated behavior on the plane $y$--$z$. It is straightforward to obtain that a solution of Eq.~\eqref{schro2d} is
\begin{eqnarray}
     f(\zeta,y,z)={\mbox{Ai}}\left(\left(\frac{a \lambda^2}{8} \right)^{1/3}\left(y+z+\frac{a}{2}\zeta^2 \right)\right)
     \exp\left(-\frac{a\lambda}{4}\zeta \left(y+z+\frac{a}{3}\zeta^2 \right) \right)\, , 
     \label{airysolutioncase2}
\end{eqnarray}
where $a$ is  the acceleration of the dynamics of this solution in the $y$--$z$ plane.
Therefore, the most general Airy solution for this second case is
\begin{equation}
    u(t,x,y,z)={\mbox{Ai}}\left(\left(\frac{a \lambda^2}{8} \right)^{1/3}\left(y+z+\frac{a}{2}\zeta^2 \right)\right)
       \exp\left(-\frac{a\lambda}{4}\zeta \left(y+z+\frac{a}{3}\zeta^2 \right) +\frac{\gamma}{c}\eta+\frac{\beta+\alpha\gamma}{c^2\lambda} \zeta\right)\, ,
    \label{asnGcase2}
\end{equation}
which is different from its counterpart \eqref{asnG} in terms of the effective acceleration of the wavepacket, and in how it propagates in the $y$--$z$ plane.
\end{widetext}

\section{Discussion}

We have proved that a general three-dimensional spatial Telegraph equation \eqref{telegraphic} always admits exact solutions in terms of Airy functions. Such solutions propagate along characteristic speed-cone-like coordinates, whereas they dynamically develop in the transverse directions. This implies that these solutions with only one Airy function do not exist for systems that have one spatial dimension, as erroneously  has been claimed in Ref.~\cite{umul}.

Differently to the usual plane-wave solutions of Eq.~\eqref{telegraphic}, the both solutions  \eqref{asnG} and \eqref{asnGcase2} have a distintive dynamical behavior. The maximum intensity lobe of the  Airy function will present different forms of acceleration (because the $\zeta^2$ dependence) in the $y-\zeta$, $z-\zeta$, or $y-z-\zeta$ planes. Besides, the position of the maximum intensity lobe will be determined by $\lambda$. Therefore, each different phenomena determine the form of the evolution of the main Airy lobe of the solution.

In addition to the above two solutions, and  depending on the values of the constants, a more general square-integrable solution \cite{lekner} can be constructed directly from solutions \eqref{asnG} and \eqref{asnGcase2}. With all these, we expect that the general solutions presented here may be used to test new phenomena in a broad set of different physical scenarios.

\section{Declarations}

{\bf Funding and/or Conflicts of interests/Competing interests:} The authors declare there are no competing interests.
{\bf Data availability statement:} This manuscript does not report data.

\begin{acknowledgements}
FAA thanks to FONDECYT grant No. 1230094 that partially supported this work.
 \end{acknowledgements}

\end{document}